\documentclass[twocolumn,showpacs,preprintnumbers,prc,superscriptaddress]{revtex4-2}
\usepackage{graphicx,color}
\usepackage{float}
\usepackage{bm}
\usepackage[pdftex,colorlinks=true, linkcolor = blue, citecolor=blue,urlcolor=blue, bookmarksnumbered=true, bookmarksopen=true]{hyperref}
\begin{document}

\title{Angular momentum removal by neutron and $\gamma$-ray emissions during fission fragment decays}

\author{I. Stetcu}
\author{A.~E. Lovell}
\author{P. Talou}
\author{T. Kawano}
\affiliation{
  Los Alamos National Laboratory, Los Alamos, New Mexico 87545, USA}

\author{S. Marin}
\author{S.~A. Pozzi}
\affiliation{Department  of  Nuclear  Engineering  and  Radiological  Sciences, University  of  Michigan,  Ann  Arbor,  Michigan  48109,  USA}

\author{A. Bulgac}
\affiliation{
Department of Physics,
  University of Washington, Seattle, Washington 98195--1560, USA}

\date{October 15, 2021}
\preprint{LA-UR-21-23498, NT@UW-21-07}

\begin{abstract}
We investigate the angular momentum removal from fission fragments (FFs) through neutron and $\gamma$-ray emission, finding that about half the neutrons are emitted
with angular momenta  $\ge 1.5\hbar$ and that the change in angular momentum after the emission of neutrons and statistical $\gamma$ rays is significant, contradicting usual assumptions.
Per fission event, in our simulations, the neutron and statistical $\gamma$-ray emissions change the spin of the fragment by 3.5 -- 5~$\hbar$, with a large standard deviation comparable to the average value. Such wide angular momentum removal distributions can hide any underlying correlations in the fission fragment initial spin values. Within our model, we reproduce data on spin measurements from discrete
transitions after neutron emissions, especially in the case of light FFs. The agreement further improves for the heavy fragments if one removes from the analysis the events that would produce isomeric states.
Finally, we show that while in our model the initial FF spins do not follow a saw-tooth like behavior observed in recent measurements, the average FF spin computed after neutron and statistical $\gamma$ emissions exhibits a shape that resembles a saw tooth. This suggests that the average FF spin  measured after statistical emissions is not necessarily connected with the scission mechanism as previously implied.
\end{abstract}
\maketitle

The angular momenta of fission fragments (FFs) have been the object of old and renewed experimental and theoretical investigations~\cite{Strutinsky:1960,Ericson1960,Huizenga:1960,Vandenbosch:1960,Nix:1965,
Rasmussen:1969,Wilhelmy:1972,Vandenbosch:1973,Moretto:1980,Dossing:1985,Moretto:1989,
Wagemans:1991,Bonneau:2007,Becker:2013,Vogt:2013a,Randrup:2014,Wilson2021,Bulgac:2021,Schunck:2020x,Randrup:2021}.
Complete theoretical modeling of the fission reaction from the formation of the compound nucleus to its splitting into two fragments and the emission of prompt neutrons and $\gamma$ rays is complicated given the time scales involved in the process and the staggering number of degrees of freedom. In practice, we usually use one type of model for the initial dynamics from compound nucleus to scission and shortly after, and another type to simulate the decay of FFs, with the goal of combining the information in order to provide a unified description of the reaction products that can later be used in various applications. 

Experimentally, the information about the FF spins is usually inferred from measurements of properties of $\gamma$ rays emitted as the FF decays towards the ground or an isomeric state~\cite{Wilhelmy:1972,Wilson2021}. As the timescales ($\sim 10^{-18}$~s) for the prompt neutron emission are governed by the nuclear interactions, the measurement of the FF properties before neutron emission is not possible. Hence, in order to extract such information, one needs to correct for the neutron emission, and for the emission of statistical $\gamma$ rays, which is always a very difficult task and possibly model dependent. Thus, a more accurate characterization of such experimental measurements of FF spins is that they provide on average only a lower bound of spin values. However, it is worth noting that while on average we find that the spin decreases with neutron and $\gamma$-ray emissions, in some rare fission events the FF spin can also increase because of angular momentum coupling.

The common lore among theorists and experimentalists alike is that the statistical neutron emission, on average, does not significantly change the FF spin. Hence, up to a small correction, the FF average spin inferred from measuring the properties of the emitted $\gamma$ rays is a good approximation of the initial FF spin. By drawing conclusions about the spin correlations at scission via measurements of the FF spin after neutron emissions, without proper simulations of angular momentum removal, the same indirect assumption is made, that the neutron evaporation from FFs does not affect such correlations. In the following, we show that this is not the case.

In this paper, we investigate the angular momentum removal by the neutrons and statistical $\gamma$ rays emitted from FFs using the Los Alamos developed codes \texttt{CGMF} \cite{CGMF-CPC} and \texttt{BeoH}, both based on the Hauser-Feshbach fission fragment decay (HF$^3$D) model~\cite{Okumura2019,Lovell:2020sdm}, and compare the results against recent experimental data. In these codes, the FFs are treated as compound nuclei that de-excite via neutron and $\gamma$-ray emission. The full competition between neutron and $\gamma$ emissions is taken into account in a Hauser-Feshbach statistical framework~\cite{HF1952}. We find that we can reproduce the trends observed in recent angular momentum data by  Wilson et. al \cite{Wilson2021} for quite a few FFs. Where the agreement is less satisfactory, we investigate possible issues. We note that while we have used the published version of the Monte Carlo code \texttt{CGMF} \cite{CGMF-CPC} for the analysis presented in this paper, we have updated the discrete-level file to include additional rotational levels that are essential to the current analysis. Thus, even though we have tried to better inform the calculations, our results can be impacted to some degree by the incomplete knowledge of the nuclear structure included in the RIPL \cite{RIPL3} database. In a recent publication, it was shown that including such high-spin states improves the description of the prompt fission neutron spectrum~\cite{Kawano2021PFNS}.

We first direct our attention to the change in spin of the FFs caused by each neutron emission. However, before getting to our results, we take a brief detour to discuss neutron emission at low energies. One of the assumptions made in a recent experimental study \cite{Wilson2021}  is that neutrons are emitted overwhelmingly as $s$ waves. Based on the known shape of the prompt fission neutrons spectrum in the center-of-mass (CM), it is reasonable to assume that most of the neutrons are emitted with energies around 1~MeV. Hence, one might expect that indeed that higher partial waves are suppressed. In the Supplemental Material \cite{supplement}, we illustrate for two representative FFs at different initial excitation energies how the competition between different relative angular momenta evolves as a function of outgoing neutron energy, finding that in a significant number of events it is more likely to emit $p$-wave and higher neutrons. One should also note that even for low-energy reactions, several partial waves can compete with $s$-waves at relatively low energies, well below 1~MeV incoming neutron energies. In the Supplemental Material we illustrate this fact by plotting the transmission coefficients for neutrons incoming on $^{95}$Sr and $^{139}$Xe targets, forming $^{96}$Sr and $^{140}$Xe compound nuclei. The $p$-wave strength function peaks in the mass 90 -- 100 region~\cite{Mughabghab2006}, often called illustratively as the nuclear Ramsauer effect~\cite{PhysRev.125.955}, which implies the importance of higher partial waves for the light FF even at low neutron energies as a consequence of quantum effects. The optical potential model incorporates such effects automatically.

\begin{table}[t]
    \centering
    \caption{The average angular momentum $\langle j_\mathrm{rm}\rangle=j_\mathrm{ini}-j_\mathrm{fin}$ (in $\hbar$ units) removed by each prompt neutron, and its standard deviation, $\Delta j_\mathrm{rm}$, for $^{235}$U($n_\mathrm{th}$,f) and $^{239}$Pu($n_\mathrm{th}$,f), $^{238}$U($n_\mathrm{1.9 MeV}$,f), and $^{252}$Cf(sf) reactions, and the percentage of neutrons that remove an angular momentum larger or equal to $\frac{3}{2}$  and $\frac{7}{2}$, respectively. Only about 25\% of the neutrons remove $\frac{1}{2}$. ``Removed" angular momentum smaller than zero means that the neutron emission increases the FF angular momentum, using the Koning-Delaroche optical potential \cite{Koning2003}.}
    \begin{tabular}{cccccc}
\hline\hline
    Reaction & $\langle j_\mathrm{rm}\rangle$ &  $\Delta j_\mathrm{rm}$ & \multicolumn{3}{c}{$j_\mathrm{rm}$} \\
    \cline{4-6}
    & &  & $ < 0$ (\%) &$\ge1.5$ (\%) & $\ge 3.5 $(\%) \\
\hline
$^{235}$U(n$_\mathrm{th}$,f) & 1.33 & 1.97  &  22.2 & 51.7 & 14.5 \\
$^{238}$U(n$_\mathrm{1.9 MeV}$,f) & 1.34 & 1.93 & 21.5 & 52.5 & 14.5 \\
$^{239}$Pu(n$_\mathrm{th}$,f) & 1.23 & 1.91  &  23.5 & 49.9 & 12.8 \\
$^{252}$Cf(sf) & 1.13  &  1.71 &    23.7  &       49.2    &       11.1 \\
\hline\hline
    \end{tabular}
    \label{tab:spinRemovedNeutron}
\end{table}

Defining the spin removed by the neutron $j_\mathrm{rm}$ as the difference between initial and final spins of the states involved in the emission of a single neutron, we find that the average spin removed is greater than $1~\hbar$, as shown in the summary table~\ref{tab:spinRemovedNeutron} for all reactions investigated in this paper. A smaller but significant fraction of the neutrons is emitted with at least 3.5~$\hbar$ angular momentum. In addition, about 25\% of the neutron emissions cause an increase of the angular momentum after emission (see Sec.~II in the Supplemental Material), in contrast with assumptions of equiprobable decrease and increase of spin by single neutron emission~\cite{Wilson2021}. These results are further illustrated in Fig.~3 in the Supplemental Material, where we show the probability to change the angular momentum in the light and heavy FFs as a function of the CM energy of the emitted neutron.  The $f$-wave neutrons do not appear to have such a high probability in Fig. 3 of the Supplemental Material because of the centrifugal barrier at those energies. The overall tendency to decrease rather than increase the angular momentum by neutron (and $\gamma$) emission is a consequence of the level densities in the residual compound nucleus, which increase with the decrease in the spin of the final state as the energy is released. We found that the average of the angular momentum removed by the first neutron is slightly smaller than the following ones, consequence again of the behavior of level densities with the excitation energies in the residual nucleus. Details are presented in Table~I of the Supplemental Material \cite{supplement}.

\begin{figure}[t]
    \centering
    \includegraphics[scale=0.38]{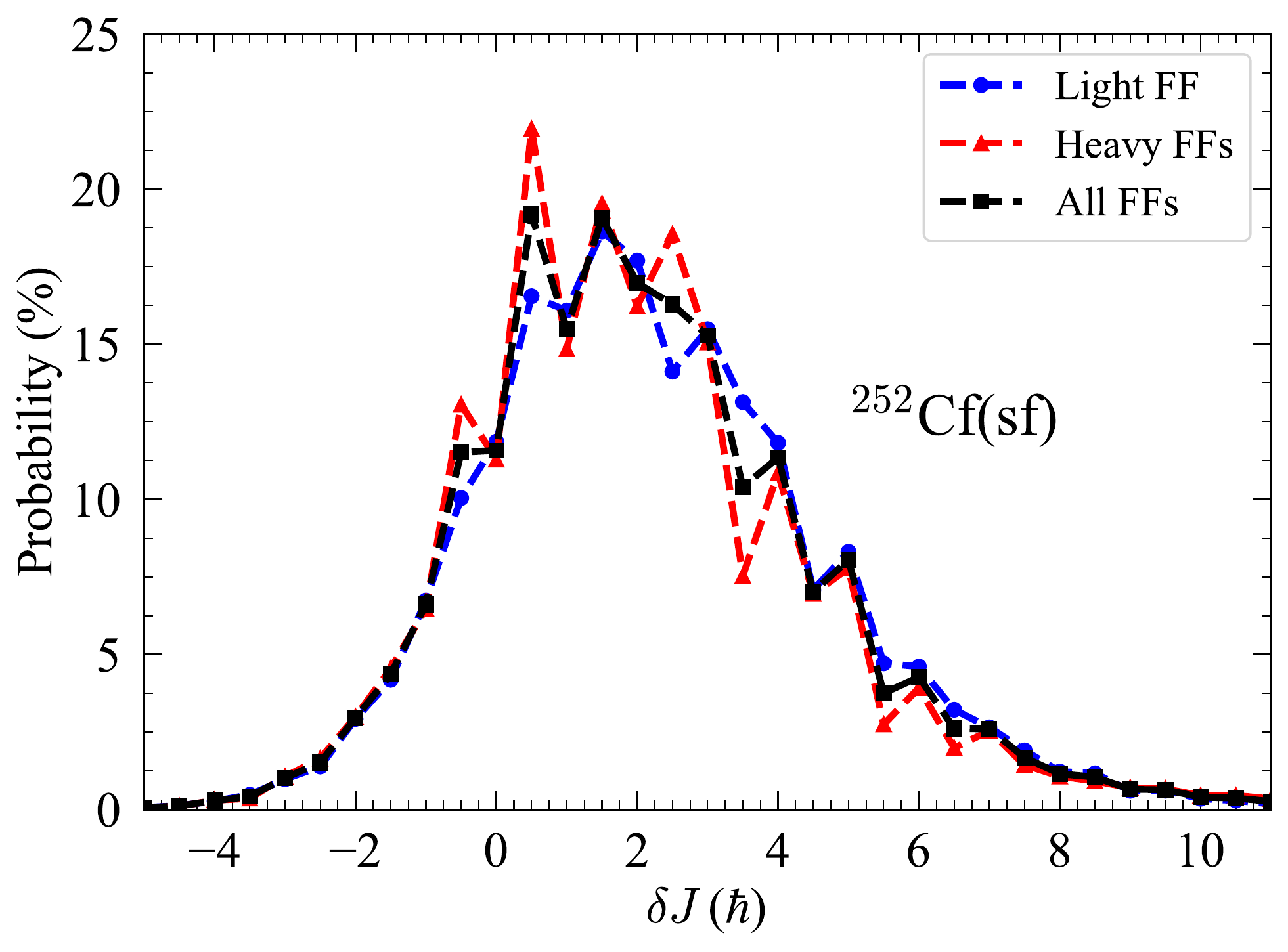}
    \caption{Distribution of the total angular momentum removed for light (filled circles), heavy (filled triangles) and all (filled squares) FFs after all prompt neutrons have been emitted. The properties of this wide distribution are listed in Table \ref{tab:totalSpin-neutrons}.}
    \label{fig:Pemission}
\end{figure}

The analysis is incomplete if we just consider the change in angular momentum after one neutron emission. It is possible that after a second neutron is emitted from the same fragment during the same fission event, the overall change in the angular momentum becomes very small. However, this is not the case in our model, as illustrated in the upper half of Table~\ref{tab:totalSpin-neutrons}, where we present the average change in the FF spin and the change in the absolute value of the FF spin after all neutrons have been emitted for the four reactions considered in this paper. The wide probability distribution for removing angular momenta in fission events by neutron emissions only is shown in Fig. \ref{fig:Pemission}. The change in spin is about 1.8~$\hbar$ and higher with a significant standard deviation of 2~$\hbar$ and higher, depending on the reaction. This is at odds with modeling in \texttt{FREYA} \cite{Randrup:2021} and the assumptions in Ref.~\cite{Wilson2021}.

\begin{table}[t]
    \caption{The angular momentum removed (in $\hbar$ units), $\delta J$, and it absolute value, $\delta |J|$, as well as their standard deviations, after all neutrons have been emitted and after all neutrons and statistical $\gamma$ rays have been emitted.}
    \label{tab:totalSpin-neutrons}
    \centering
    \begin{tabular}{ccccc}
    \hline\hline
    Reaction & $\delta J$ & $\Delta_{\delta J}$ & $\delta |J|$ & $\Delta_{\delta |J|}$ \\
    \hline
   \multicolumn{5}{c}{ Neutron emission only} \\
    $^{235}$U(n$_\mathrm{th}$,f) &  1.84 & 2.35 & 2.19 & 2.04 \\
    $^{238}$U(n$_\mathrm{1.9 MeV}$,f) & 1.98 & 2.41 & 2.30 & 2.10\\
    $^{239}$Pu(n$_\mathrm{th}$,f) & 1.93 & 2.44 & 2.29 & 2.11 \\
    $^{252}$Cf(sf)& 2.20 & 2.51 & 2.56 & 2.15 \\    
    \hline
   \multicolumn{5}{c}{Neutron and statistical $\gamma$-ray emission} \\
    $^{235}$U(n$_\mathrm{th}$,f) & 3.54 & 3.78 & 3.82 & 3.50  \\
    $^{238}$U(n$_\mathrm{1.9 MeV}$,f) & 4.34 & 4.42 & 4.58 & 4.17\\
    $^{239}$Pu(n$_\mathrm{th}$,f) & 3.92 & 4.13 & 4.20 & 3.84\\
    $^{252}$Cf(sf) & 4.98 & 4.93 & 5.23 & 4.66 \\
    \hline\hline 
    \end{tabular}
\end{table}


For a complete analysis, we also need to consider the $\gamma$-ray emission. The lower part of Table~\ref{tab:totalSpin-neutrons} shows the average and standard deviation of the change in spin and absolute value of the spin after both neutrons and statistical $\gamma$ rays have been emitted, \textit{i.e.,} until the decay reaches the first discrete transition in the FF. Overall, the average angular momentum removed is rather large, between 3.5~$\hbar$ and 5~$\hbar$, depending on the reaction, with the standard deviations comparable with the average. We expect the characteristics of these distributions to be robust and survive a more thorough sensitivity analysis  \cite{Okumura2019}.

\begin{figure}[t]
    \centering
    \includegraphics[scale=0.42]{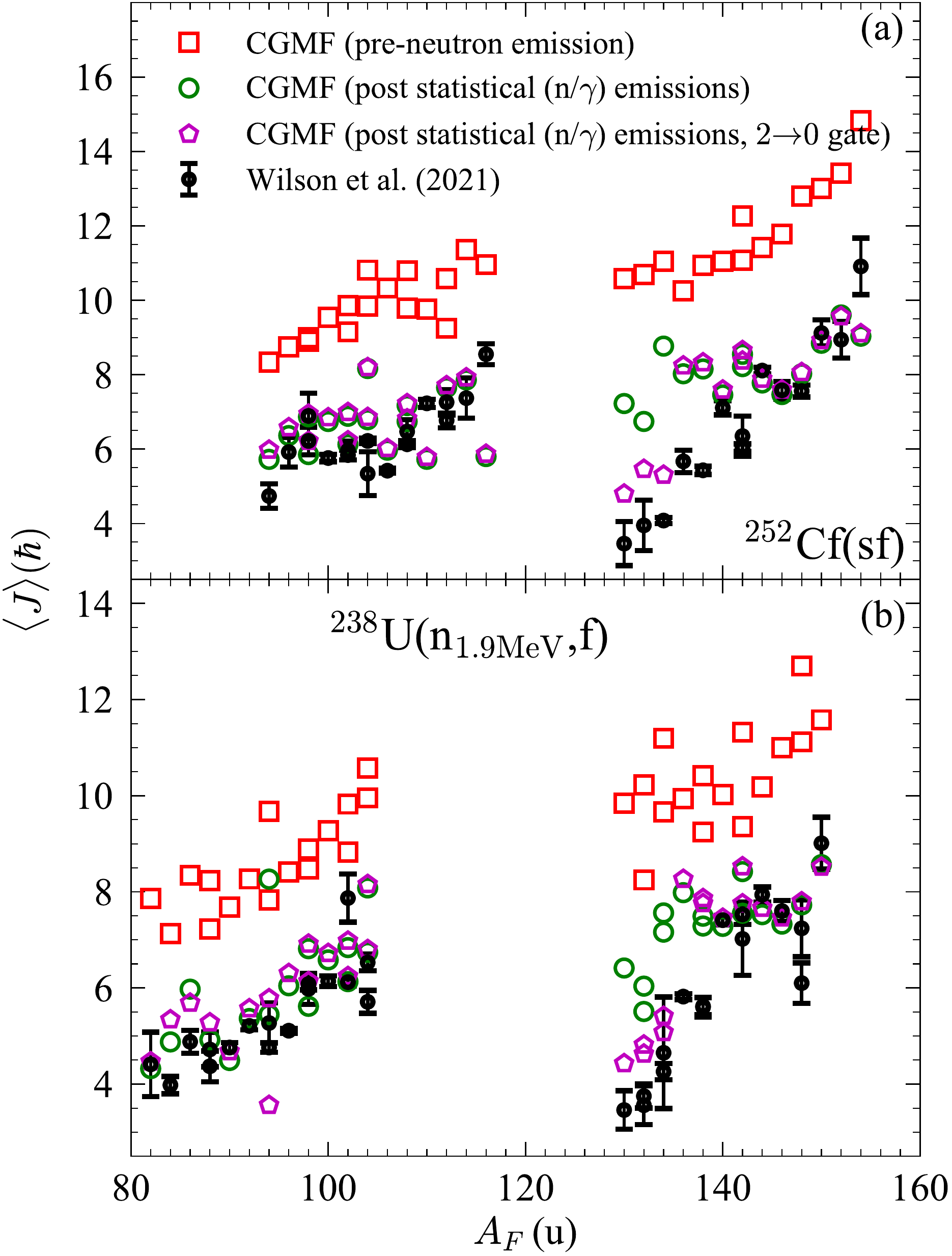}
    \caption{The average angular momentum of the primary FFs producing select residuals with mass $A_F$ for (a) $^{252}$Cf(sf) and (b) $^{238}$U(n$_\mathrm{1.9 MeV}$,f). We compare the data by Wilson et. al.~\cite{Wilson2021} with \texttt{CGMF} simulations. Three sets of results are presented for \texttt{CGMF}: one obtained by averaging the initial spin of all the FFs producing a targeted residual (squares), the second one by applying Eq. (\ref{eq:avJ}) (circles), and the third one similar to the second one, with an additional gate to allow only events with $2^+\to0^+$ as the last $\gamma$ transition in the cascade. For a more meaningful comparison, we have added 1~$\hbar$ to the latter two \texttt{CGMF} results, which is the correction applied in the experiment~\cite{Wilson2021} to account for the change in angular momentum during the statistical neutron and $\gamma$ emissions.}
    \label{fig:avAngMom-Discrete}
\end{figure}

In order to make a one-to-one comparison between experiment and theory, we need to cast the main procedure of extracting the spin within the language of our Monte Carlo implementation. The side feeding used in the measurements is defined as the difference between the intensity of transitions going into a level and the intensity of transitions going out of the level. In an event-by-event theoretical framework, the side feeding is non-zero only at a long-lived isomer, at the ground state, and at the highest energy levels where the discrete-to-discrete transitions start in a decay event. 
According to Ref.~\cite{Wilson2021} one can neglect the isomeric states as their contribution is small, even though we found otherwise as shown below. The simulations are complete as long as the information about the discrete level properties including spin assignment and branching ratios are complete. If we denote by $\tilde J^D$ the spin of the (highly excited) discrete level where the first discrete-to-discrete $\gamma$-ray transition occurs, the average FF angular momentum after neutron and statistical $\gamma$-ray emissions is

\begin{equation}
    \langle J\rangle=\tilde{C}\sum_{i}\tilde N_i \tilde J^D_i,
    \label{eq:avJ}
\end{equation}
where the sum runs over the all events producing the chosen FF after neutron emission and $\tilde N_i$ is the number of times the particular state $i$, with spin $\tilde J^D_i$, is reached first during the simulation ($\tilde{C}$ =$1/\sum_i \tilde N_i$). Because our average spins are calculated at the first discrete state, to better compare with the data of Wilson et. al.~\cite{Wilson2021}, we add 1$\hbar$ to the values given by Eq.~(\ref{eq:avJ}), which is the value they claim is reasonable to correct for statistical neutron and $\gamma$-ray emission. In Ref.~\cite{Wilhelmy:1972}, a statistical model has been employed to account for both neutron and statistical $\gamma$-ray angular momentum removal, but no details on the size of the corrections are given.  With this definition, we find that our results, based on Eq.~(\ref{eq:avJ}) and marked by green circles in Fig.~\ref{fig:avAngMom-Discrete}, are in reasonable agreement with the experimental data, especially for the light fragments. 

We have also calculated the average spin of the FFs by considering properties before neutron emission. Given that the spin measurement is made after neutron emission, a targeted mass $A_F$ will be produced by events with FF mass $A_0=A_F+n$, with $n\geq0$. Hence, the initial average spin for measured FF with mass $A_F$, marked in Fig.~\ref{fig:avAngMom-Discrete} by red squares, is obtained by selecting all the events with $A_0$ that produce $A_F$ after neutron emission.
This average is significantly higher than the one based on Eq.~(\ref{eq:avJ}), marked by green circles in Fig.~\ref{fig:avAngMom-Discrete}, and the main reason is that the angular momentum removal by statistical emissions is underestimated if a constant 1~$\hbar$ correction is applied, as also noted in Ref. \cite{Schunck:2020x}. In Table~\ref{tab:totalSpin-neutrons} we report much higher average values for the removed spin. However, as illustrated with green circles, the simulated average spin values do show similar trends (except for nuclei in the neighborhood of closed shell) as in the experimental data~\cite{Wilson2021}. We note that because $^{130,132}$Sn and $^{134}$Te have long-lived isomeric states, some Monte-Carlo cascades do not end in a $2^+\to0^+$ transition as these very high-spin states do not have time to decay in the time coincidence window of 10~ns that we impose and is usual in these types of experiments. Because experimentally one looks for the rotational band transitions, we have eliminated all the events that do not end in a $2^+\to0^+$ $\gamma$-ray transition. This leaves most of the results in Fig.~\ref{fig:avAngMom-Discrete} unchanged, with the exception of $^{130,132}$Sn and $^{134}$Te which now better reproduce the data. This result also shows that while our initial FF average spin distribution that should be produced at scission exhibits no saw-tooth behavior, the resulting spins after statistical neutron and $\gamma$ emission can have that behavior as a consequence of spin removal from the compound nuclei. Thus, this explanation is in contrast with the geometrical interpretation based on an \textit{ad hoc} parameterization ${\cal I}'(A_f)=0.2{\cal I}_\mathrm{rig}(A_f;0)+2({\cal I}_\mathrm{rig}(A_f;\epsilon_\mathrm{sc})-{\cal I}_\mathrm{rig}(A_f;0))$~\cite{Randrup:2021}, the discrepancy probably lies in the different modeling of neutron and statistical $\gamma$ emissions, including the  treatment of the angular emission by neutrons (classical in Ref. \cite{Randrup:2021} vs. fully quantum mechanical in this work).

Even before the data in Ref.~\cite{Wilson2021} were published, there was evidence that in our statistical model the spins of the heavy FFs are somewhat overestimated, especially for FFs around closed shells. Microscopic calculations predict that the average light FF has a larger angular momentum than its heavy counterpart~\cite{Bulgac:2021,Schunck:2020x}, especially in the neighbourhood of closed-shell configurations. However, even though the authors of Ref.~\cite{Wilson2021} expressed confidence of their uncertainties, their method is also based heavily off of yrast transitions. In the Sn region, the nuclei are either spherical or weakly deformed, and hence the rotational band is not very well defined. We have obtained average spins results for the other reactions studied in this paper, but we present the results in the Online Supplemental Material, since no experimental data are available in these cases.

With the large change in angular momentum that occurs during the neutron and statistical $\gamma$-ray emissions, it could be difficult to extract correlations between nascent FFs from measurement of spins after neutrons emission. As noted in Ref.~\cite{Randrup:2021}, even if the mechanism generates fragments with strongly aligned spins, the resulting angular momenta appear largely uncorrelated. In the HF$^3$D model, the average angular momenta are highly correlated, since the cutoff parameter, which determines the spin distribution, depends on the excitation energy in each fragment~\cite{Becker:2013,Stetcu:2014prc}, and the excitation energies in turn are correlated via energy conservation. However, because of the significant width in excitation energy distribution in each FF, the correlations in spin values are significantly diluted (correlation coefficients $\sim \pm0.01$). Hence, in the HF$^3$D model the spin fragments appear uncorrelated even though the mechanism that generates the spins should produce highly correlated average angular momenta. Even when looking at the initial spin of the FFs, we see no correlation between the heavy and the light spins.

Finally, there are other types of correlations that would not be accessible by only measuring properties of FFs after neutron (and part of $\gamma$) emission, in particular the bending and twisting modes theoretically conjectured in the 1960's~\cite{Strutinsky:1960,Nix:1965} and recently in microscopic calculations~\cite{Bulgac:2021}. It is also likely that the geometrical correlations found in Ref.~\cite{Bulgac:2021,Bulgac:2021b}, namely the FF spins are generated prior to any emission at angles slightly higher than $\pi/2$, may also translate into angular correlations between emitted particles.  Since the equilibrated FFs emerge typically elongated along the fission axis~\cite{Bulgac:2019d,Bulgac:2020}, in a simplified model the neutron emission will be from the FF tips, where the suppression due to the centrifugal barrier is minimal. Thus, one might expect an enhancement of such angular correlations. 


We have investigated the angular momentum change caused by the neutron and $\gamma$-ray emission from equilibrated FFs during the evolution toward stable states, before any $\beta$ decay. We have shown that in the framework of the HF$^3$D model, the neutrons and statistical $\gamma$ rays remove a significant amount of angular momentum. Inevitably, in any model that simulates such a complex phenomenon, which involves several nuclei far from stability, some of the systematics extracted from data involving stable or long-lived isotopes will turn out to be less reliable. Microscopic calculations can help, but currently they are neither precise nor detailed enough to be directly used in calculations without calibrations. 
Experimental data can help in calibrating the phenomenological models, and we found that trends in 
recent~\cite{Wilson2021}  data agree reasonably well with our approach based on the HF$^3$D model. However, because the FF spins immediately after scission cannot be directly inferred from measurements before neutron emission, some of the interpretations could be subject to model dependence, if neutron emission corrections are treated in a particular approach. This statement is not only valid for angular momentum measurements, but also for other physical observables that need to be corrected for neutron and $\gamma$ emissions. In particular we have shown that details in computing the angular momentum removed by statistical decays can produce a saw-tooth-like behavior that does not come from the mechanism of generating the angular momentum at scission, and some aspects can be enhanced by the presence of isomeric states.


\section*{Acknowledgments}

We thank N. Fotiades, M. Develin and G. Rusev for useful discussions. This work was carried out under the auspices of the
National Nuclear Security Administration of the U.S. Department of
Energy at Los Alamos National Laboratory under Contract
No. 89233218CNA000001. We acknowledge partial support by the Office of Defense Nuclear Nonproliferation Research \& Development and by the Nuclear Criticality Safety Program, funded and managed by the National Nuclear Security Administration for the Department of Energy. This work was funded in-part by the Consortium for Monitoring, Technology, and Verification under Department of Energy National Nuclear Security Administration award number DE-NA0003920.
AB was supported by U.S. Department of Energy,
Office of Science, Grant No. DE-FG02-97ER41014 and in part by NNSA
cooperative Agreement DE-NA0003841.

\bibliography{references}

\newpage 

\begin{widetext}

\section*{Supplemental Online material}

\setcounter{figure}{0}
\setcounter{table}{0}
\subsection{Neutron emission probabilities}

The neutron emission from fission fragments (FFs) is modeled within the statistical theory of Hauser and Feshbach \cite{HF1952}.  As an example, we show in Fig. \ref{fig:DecayProb} the decay probability of neutron emission for various partial waves as a function of the emitted neutron energy for two representative FFs, $^{96}$Sr and $^{140}$Xe, each with 10 and 20 MeV excitation energy available for neutron emission. As expected, this figure shows that the competition between emission with different relative angular momenta depends on the excitation energy available for neutron emission. 

\begin{figure*}[b]
    \centering
    \includegraphics[scale=0.45]{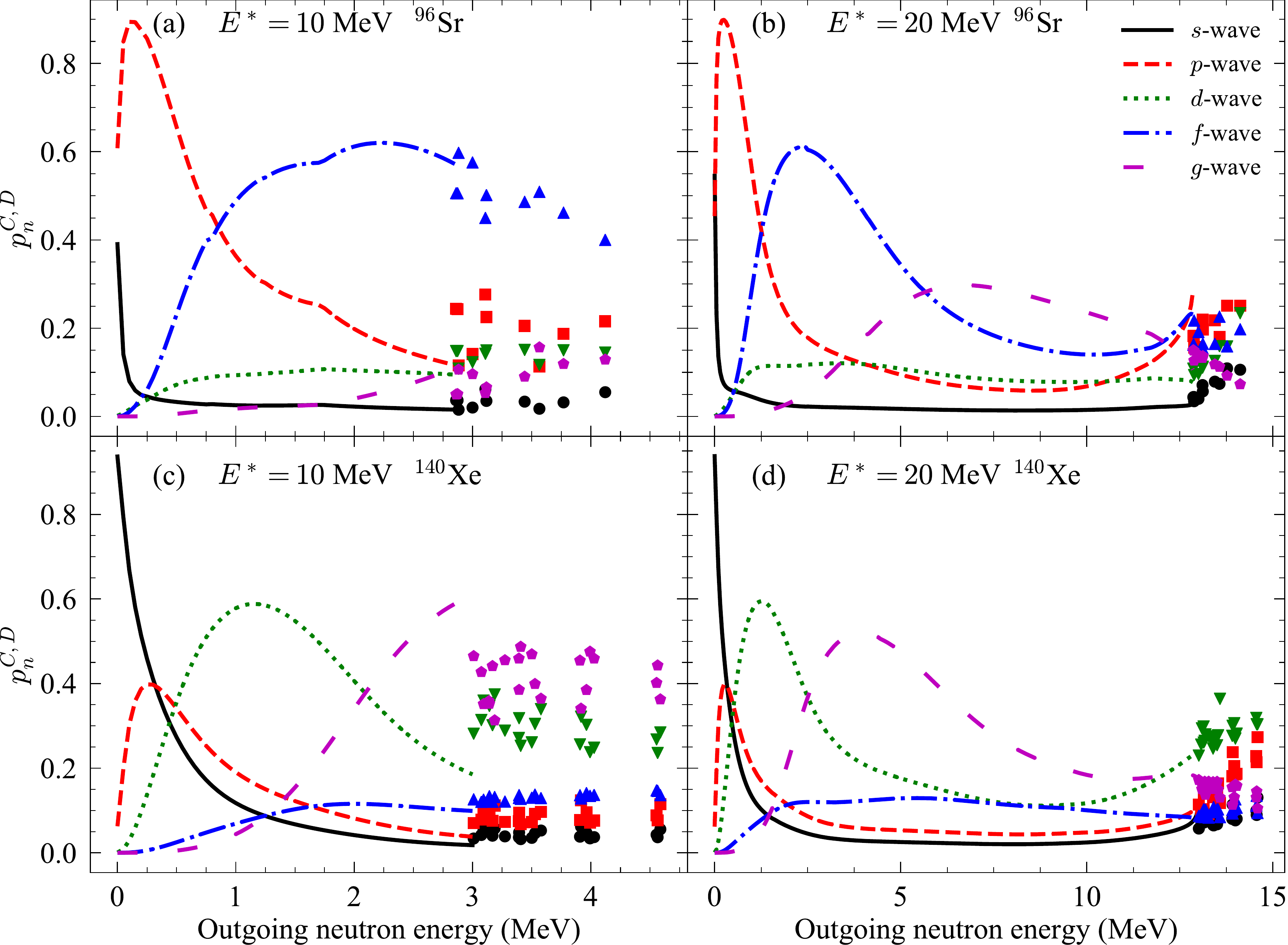}
    \caption{Decay probability by neutron emission for (a,b) $^{96}$Sr and (c,d) $^{140}$Xe as a function of the outgoing neutron energy, for different neutron relative angular momenta. For these examples, the total excitation energy available for decay has been set to 10 MeV. The lines represent the probability of decay into continuum states $p_n^C$, while discrete points (circles for $s$-wave, squares for $p$-wave, down triangles for $d$-wave, up triangles for $f$-wave, and pentagons for $g$-wave) mark the transition probability into discrete states  $p_n^D$ of $^{95}$Sr and $^{139}$Xe, respectively. }
    \label{fig:DecayProb}
\end{figure*}

\begin{figure}[h]
    \centering
    \includegraphics[scale=0.45]{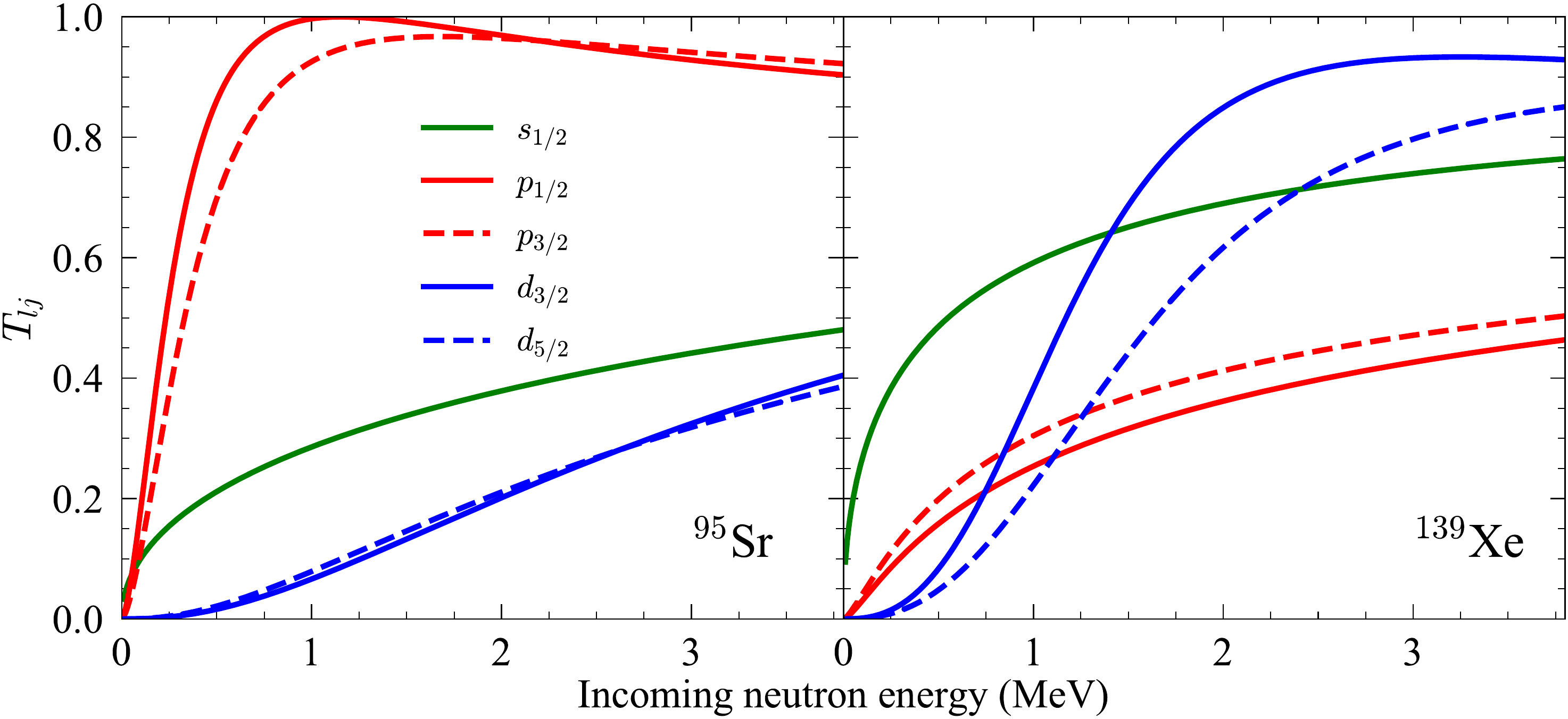}
    \caption{The transmission coefficients for neutrons on $^{95}$Sr (left panel) and $^{139}$Xe (right panel) as a function of incident neutron energy. We show $l=0, 1, 2$ partial waves. Note that when the transition matrix element is calculated, an additional $2j+1$ has to be included, which increases the contribution of higher partial waves at energies below 1~MeV even for the $^{139}$Xe target.}
    \label{fig:TransitionCoeffs}
\end{figure}

In the lighter fragment, $^{96}$Sr, the transmission coefficients for relative neutron spins $l>0$ are comparable and often larger than for $s$-wave emission throughout the entire outgoing neutron energy range. Even for $^{140}$Xe, where the $s$-wave emission dominates at very low energies, its probability becomes comparable with $p$- and $d$-wave emission around 200 keV, for the 10 MeV excitation energy scenario. In Fig. \ref{fig:En-Spin-corr} we show the probability to change the angular momentum in the light and heavy FFs as a function of the CM energy of the emitted neutron.

The neutron emission is the inverse process of the incident neutron impinging on a target, and the transition matrix elements are related, based on the transmission coefficients into the compound system. Hence, to show that even at energies below 1 MeV several partial waves can contribute, we plot in Fig. \ref{fig:TransitionCoeffs} the transmission coefficients for incoming neutrons on $^{95}$Sr and $^{139}$Xe. In the case of $^{95}$Sr, the $p$-wave ($l=1$)dominates at a few tens of keV. Even for the $^{139}$Xe target, the $p$-wave becomes significant at around 1~MeV incident energy.


\subsection{Statistical properties of FF spin removed by neutrons and statistical $\gamma$ rays}

We first consider in this section, details of the angular momentum removed by the FFs. In Table \ref{tab:details} we show average angular momentum removed by each neutrons if we take into account all the neutrons, only the first, the second or the third. We look at all, light anf heavy FF separately. The first neutron out tends to remove the least angular momentum, additional neutrons removing increasingly more angular momentum, as they reach lower excitation energies in the residual FFs. 

\begin{table}[h]
    \centering
    \caption{Details of the angular removal by the neutron emission from FFs. We show averages for each neutrons emitted by all, light and heavy fragments, as well as a breakdown of the angular momentum removed by the first, second and third neutrons from FFs. All momenta are in units of $\hbar$.}
    \begin{tabular}{ccccccccccccccc}
\hline\hline 
         Reaction &  \multicolumn{4}{c}{All FFs} & & \multicolumn{4}{c}{Light FFs} & & \multicolumn{4}{c}{Heavy FFs}\\
         \cline{2-5}
         \cline{7-10}
         \cline{12-15}
         & all & $1^\mathrm{st}$ &$2^\mathrm{nd}$ & $3^\mathrm{rd}$  
         & & all & $1^\mathrm{st}$ &$2^\mathrm{nd}$ & $3^\mathrm{rd}$
         &  & all & $1^\mathrm{st}$ &$2^\mathrm{nd}$ & $3^\mathrm{rd}$\\
           
    \hline
    $^{235}$U(n$_\mathrm{th}$,f) & 1.33 & 1.27 & 1.47 & 1.57 & & 1.22 & 1.14 & 1.38 & 1.54 & & 1.48 & 1.43 & 1.65 & 1.69 \\
    $^{238}$U(n$_\mathrm{1.9 MeV}$,f) & 1.34& 1.30 & 1.44 & 1.30 & &
    1.22 & 1.19 & 1.29 & 1.24 & &
    1.51 & 1.43 & 1.73 & 1.60\\
    $^{239}$Pu(n$_\mathrm{th}$,f) & 1.23 & 1.16 & 1.34 & 1.38 & & 1.16 &
    1.14 & 1.21 & 1.16 & & 1.48 & 1.19& 1.48& 1.52\\
    $^{252}$Cf(sf) & 1.13 & 1.07& 1.21 & 1.19 & & 1.07 & 1.03 & 1.09 & 1.12 &
    & 1.22& 1.12 & 1.37 & 1.37\\
    \hline \hline
    \end{tabular}
    \label{tab:details}
\end{table}

\begin{figure}
    \centering
     \includegraphics[scale=0.40]{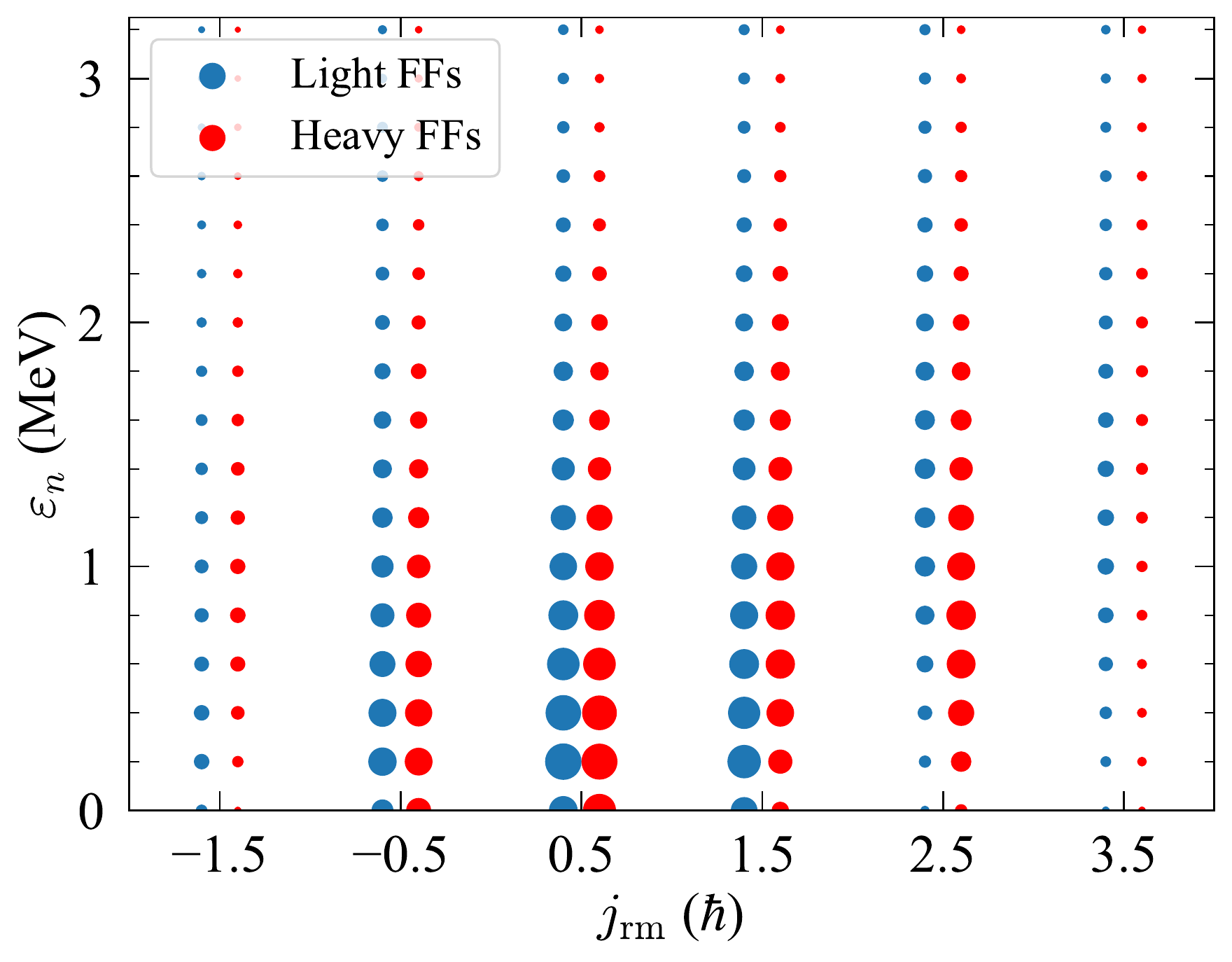}
    \caption{Correlations between the CM energy of the emitted neutrons and the change in the FF angular momentum for the $^{252}$Cf(sf) reaction, for the light and heavy FFs. We added a small negative and positive offset to distinguish between light and heavy FFs, respectively. The sizes of the dots are proportional to the number of events produced.}
    \label{fig:En-Spin-corr}
\end{figure}

In Table \ref{tab:totalSpin-neutrons-and-gammas} we report the average angular momentum removed (and its absolute value) , as well as the corresponding standard deviations, after neutron emission and after both neutron and statistical $\gamma$-ray emission.  We also show the percentage of FFs with a removal of spin in selected intervals, with about 35\% changing by more than $2\hbar$ for only the neutron emission. In addition, more than 14\% of the events result in an increase in FF spin after all neutrons have been emitted.  The distribution of the total angular momentum removed by neutron emission is shown in Fig. 1(b) in the main manuscript.  While the highest number of fission events changes the FF spins between 0 and $2\hbar$, there is a non negligible number of fission events that change the FF spins by $2\hbar$ or more. On the other hand, the number of events that removes more than 2$\hbar$ after neutrons and statistical $\gamma$ emission is 55\% and higher.

\begin{table}[b]
    \caption{Total angular momentum removed $\delta J$ and the absolute value of the angular momentum removed $\delta |J|$, with the standard deviation in parentheses. For each reaction, we also show the percentage of fission events in which the change in the total angular momentum changes by negative values (\textit{i.e.}, spin increases), between 0 and 1, between 1 and 2 and greater than 2. The spins are expressed in units of $\hbar$. In the upper part we present the spin removal after all neutrons have been emitted from FFs, while in the lower part we show the spin removal after both neutrons and statistical $\gamma$ rays have been emitted.}
    \label{tab:totalSpin-neutrons-and-gammas}
    \centering
    \begin{tabular}{ccccccc}
    \hline\hline
    Reaction & $\delta J (\Delta_{\delta J})$ & $\delta |J| (\Delta_{\delta |J|})$ &
    $\delta J<0$ (\%) & $0\le \delta J \le 1$ (\%)& $1 < \delta J \le 2$ (\%) &
    $2< \delta J$ (\%)\\
    \hline
    \multicolumn{7}{c}{Neutron emission only} \\
    $^{235}$U(n$_\mathrm{th}$,f) &  1.84(2.35) & 2.19(2.04) & 17.0 & 25.4 & 20.7 & 36.8 \\
    $^{238}$U(n$_\mathrm{1.9MeV}$,f) &  1.98(2.41) & 2.30(2.10) & 15.6 & 24.5 & 20.3 & 39.6 \\
    $^{239}$Pu(n$_\mathrm{th}$,f) & 1.93(2.44) & 2.29(2.11) & 16.5 & 25.1 & 20.0 & 38.4\\
    $^{252}$Cf(sf)& 2.20(2.51) & 2.56(2.15) & 14.5 & 23.1 & 18.0 & 44.4 \\
    \hline
    \multicolumn{7}{c}{All statistical emissions (neutrons and $\gamma$ rays)} \\
    $^{235}$U(n$_\mathrm{th}$,f) &  3.54(3.78) & 3.82(3.50) & 11.9 & 17.8 & 13.6 & 56.8 \\
    $^{238}$U(n$_\mathrm{1.9MeV}$,f) &  4.34(4.42) & 4.58(4.17) & 10.3 & 15.5 & 11.9 & 62.3 \\
    $^{239}$Pu(n$_\mathrm{th}$,f) &3.92(4.13) & 4.20(3.84) &  11.3 & 16.0 & 12.9 & 59.8 \\
    $^{252}$Cf(sf) & 4.98(4.93) & 5.23(4.66) & 9.4 & 13.0 &  10.6 & 67.0 \\
    \hline\hline 
    \end{tabular}
\end{table} 

\subsection{Other reactions}

While data are not available for the fission induced by thermal neutrons on $^{235}$U and $^{239}$Pu, for completeness we show in Fig. \ref{fig:avAngMom-Discrete-PuU} out results for these reactions. They are rather similar to the ones shown in Fig. 2 of the main text.

\begin{figure}
    \centering
    \includegraphics[scale=0.44]{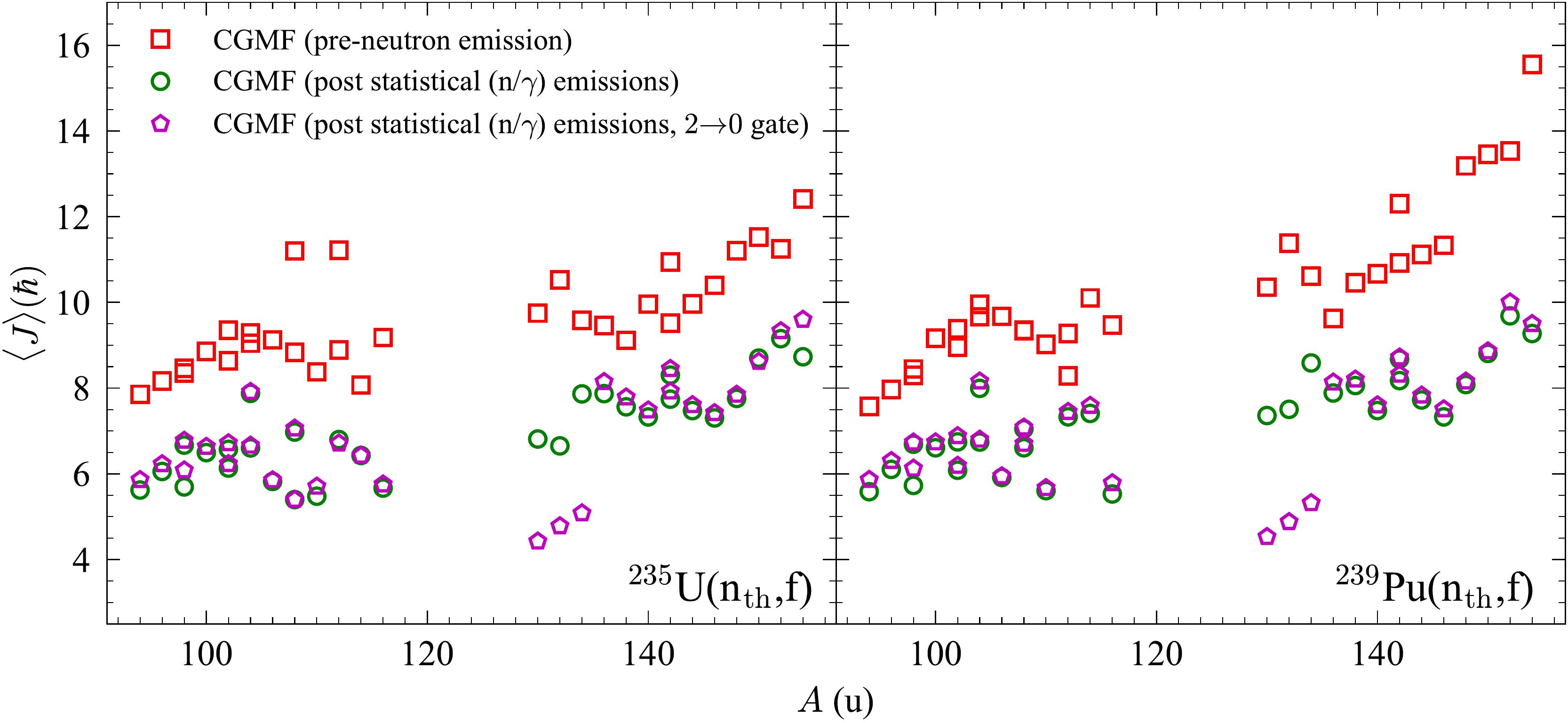}
    \caption{The average angular momentum of the primary FFs producing select residuals with mass $A$ for the induced fission by thermal neutrons on $^{235}$U (left panel) and $^{239}$Pu (right panel)  targets. Three sets of results are presented for \texttt{CGMF}: one obtained by averaging the initial spin of all the FFs producing a targeted residual (squares), the second one by applying Eq. (1) in the main paper (circles), and the third one similar to the second one, with an additional gate to allow only events with $2^+\to0^+$ as the last $\gamma$ transition in the cascade. We have kept the convention of adding $1\hbar$ to our results, as noted in the main text.}
    \label{fig:avAngMom-Discrete-PuU}
\end{figure}

\end{widetext}

\end{document}